\documentclass[12pt,preprint,flushrt]{aastex}
\usepackage[table]{xcolor}
\usepackage{booktabs}
\usepackage{tabularx}

\begin{document}

\title{A Multiwavelength Study of the Relativistic Tidal Disruption Candidate {\it Swift} J2058.4+0516 at Late Times}
\author{Dheeraj R.~Pasham$^{1,2}$, S.~Bradley Cenko$^{1,2}$,
Andrew J.~Levan$^{3}$,
Geoffrey C. Bower$^{4}$,
Assaf Horesh$^{5}$,
Gregory C.~Brown$^{3}$,
Stephen Dolan$^{6}$,
Klaas Wiersema$^{7}$,
Alexei V.~Filippenko$^{8}$,
Andrew S.~Fruchter$^{9}$,
Jochen Greiner$^{10}$,
Rebekah A.~Hounsell$^{9}$,
Paul T.~O'Brien$^{7}$,
Kim L.~Page$^{7}$,
Arne Rau$^{10}$,
and Nial R.~Tanvir$^{7}$
}
\altaffiltext{1}{Code 661, Astrophysics
Science Division, NASA's Goddard Space Flight Center, Greenbelt, MD
20771, USA}
\altaffiltext{2}{Joint Space-Science Institute, University of Maryland, College
Park, MD 20742, USA}
\altaffiltext{3}{Department of Physics, University of Warwick, Coventry, CV4 7AL, UK}
\altaffiltext{4}{Academia Sinica Institute of Astronomy and Astrophysics, 645 N. 
A\'ohoku Place, Hilo, HI 96720, USA}
\altaffiltext{5}{Benoziyo Center for Astrophysics, Faculty of Physics, The Weizmann 
Institute of Science, Rehovot 76100, Israel}
\altaffiltext{6}{Oxford Astrophysics, Denys Wilkinson Building, Keble Road, Oxford, 
OX1 3RH, UK}
\altaffiltext{7}{Department of Physics and Astronomy, University of Leicester, 
University Road, Leicester, LE1 7RH, UK}
\altaffiltext{8}{Department of Astronomy, University of California, Berkeley, CA 
94720-3411, USA}
\altaffiltext{9}{Space Telescope Science Institute, 3700 San Martin Drive, Baltimore, 
MD 21218, USA}
\altaffiltext{10}{Max-Planck-Institut f{\"u}r Extraterrestrische Physik, 
Giessenbachstra{\ss}e 1, D-85748, Garching, Germany}

\email{Email: dheerajrangareddy.pasham@nasa.gov}

\begin{abstract}
{\it Swift} J2058.4+0516 (Sw~J2058+05, hereafter) has been suggested as the second member 
(after Sw~J1644+57) of the rare class of tidal disruption events accompanied by relativistic 
ejecta. Here we report a multiwavelength (X-ray, ultraviolet/optical/infrared, radio) analysis 
of Sw~J2058+05 from 3 months to 3\,yr post-discovery in order to study its properties and 
compare its behavior with that of Sw~J1644+57. Our main results are as follows. (1) The 
long-term X-ray light curve of Sw~J2058+05 shows a remarkably similar trend to that of Sw~J1644+57. 
After a prolonged power-law decay, the X-ray flux drops off rapidly by a factor of $\ga 160$ 
within a span of $\Delta$$t$/$t$ $\le$ 0.95. Associating this sudden decline with the transition 
from super-Eddington to sub-Eddington accretion, we estimate the black hole mass to be in the 
range of $10^{4-6}$\,M$_{\odot}$. (2) We detect rapid ($\la 500$\,s) X-ray variability before 
the dropoff, suggesting that, even at late times, the X-rays originate from close to the black 
hole (ruling out a forward-shock origin). (3) We confirm using {\it HST} and VLBA astrometry 
that the location of the source coincides with the galaxy's center to within $\la 400$\,pc (in 
projection). (4) We modeled Sw~J2058+05's ultraviolet/optical/infrared spectral energy 
distribution with a single-temperature blackbody and find that while the radius remains more or 
less constant at a value of $63.4 \pm 4.5$ AU ($\sim 10^{15}$\,cm) at all times during the 
outburst, the blackbody temperature drops significantly from $\sim$\,30,000\,K at early times 
to a value of $\sim$\,15,000\,K at late times (before the X-ray dropoff). Our results strengthen 
Sw~J2058+05's interpretation as a tidal disruption event similar to Sw~J1644+57. For such systems, 
we suggest the rapid X-ray dropoff as a diagnostic for black hole mass.
\end{abstract}


\newpage

\section{Introduction \& Background}
When a star orbits close to a massive black hole ($M_{\rm BH} \ga 10^{4}$\,M$_{\odot}$) such 
that its periastron distance is $\lesssim R_{\ast}(M_{\rm BH}/m_{\ast})^{1/3}$ (where $R_{\ast}$ and $m_{\ast}$ are the radius and the mass of the star, respectively), it 
will be disrupted and cause what is commonly referred to as a tidal disruption event (Hill 
1975; Rees 1988). A fraction (roughly 50\%) of the stellar debris escapes while the rest is 
put in a highly eccentric orbit around the black hole, triggering the accretion process (e.g., 
Evans \& Kochanek 1989; Lodato \& Rossi 2011). These events are unique in the sense that they 
provide a one-time opportunity to study the onset of accretion and the formation of accretion 
disks and jets, which are currently only poorly understood.

For disrupting black holes with masses $\la 4 \times 10^{7}$\,M$_{\odot}$, the initial 
accretion rate can exceed their Eddington limit by a factor of a few tens (e.g., Giannios 
\& Metzger 2011). Numerical studies suggest that such high accretion rates should produce 
outflows/jets driven by strong radiative pressure forces (e.g., Ohsuga et al. 
2005). Although the precise jet launching mechanism is still highly debated (see 
Tchekhovskoy et al. 2010, and references therein), we know from X-ray and radio observations 
of black hole binaries and active galactic nuclei that jets and accretion are mutually dependent (e.g., Merloni et 
al. 2003; Falcke et al. 2004; Plotkin et al. 2012). Therefore, accretion initiated by the 
tidal disruption of a star is anticipated to be a natural site for producing jets.

Given that the black hole jet directions, are uniformly distributed 
over the sky, most of the jetted events will be offset from our line of sight owing to 
collimation. Theoretical studies suggest that off-axis relativistic jets, although initially unobservable 
because of Doppler beaming, should be detectable after a few years when the ejecta slow down to 
mildly relativistic speeds (Giannios \& Metzger 2011). But recent radio follow-up studies 
of tidal disruption events (TDEs) spanning 1--22\,yr after the initial disruption have 
detected radio emission from only $\la 17$\% of the sample (see Tables 1 \& 2 of van 
Velzen et al. 2013), suggesting that maybe only a specific subset of events --- those requiring 
special conditions --- produce relativistic jets (Bower et al. 2013; van Velzen et al. 2013).

{\it Swift} J164449.3+573451 (Sw~J1644+57, hereafter) is the first and the best-studied 
relativistic TDE (one accompanied by a relativistic outflow; e.g., Levan et al. 2011; Bloom 
et al. 2011; Burrows et al. 2011; Zauderer et al. 2011, 2013). The main observed properties 
of this source are as follows. (1) Long-lived ($\Delta t \approx 1$\,yr), luminous ($L_{\rm X,iso} \approx 10^{47}$\,erg\,s$^{-1}$), rapidly variable X-ray emission with a power-law secular 
decline; (2) self-absorbed radio emission indicative of relativistic ejecta; (3) location 
consistent with the nucleus of a redshift $z = 0.354$, compact, mildly star-forming 
galaxy; and (4) significant ($\sim 7$\%) near-infrared (NIR) polarization, strongly favoring an
on-axis viewing angle (Wiersema et al. 2012).  Observations at late times ($\ga 100$\,d) 
have both reinforced and complicated this picture.

The overall trend of Sw~J1644+57's X-ray light curve, neglecting the short-timescale 
variability, can be described by a more or less constant plateau stage in the first 10\,d 
(rest frame)$\footnote{All of the durations quoted in this paper will be accompanied by a 
qualifier indicating whether they were calculated in the rest frame or the observer frame. For 
instances where a qualifier is not given, it should be assumed that the values are in the 
observer frame.}$ followed by a power-law decline with an index consistent with both 
$-5/3$ and $-2.2$, corresponding to a complete and a partial disruption of the star$\footnote{The 
disruption is partial if the mass lost by the star is $\la 50$\%, while it is referred to 
as complete if the star loses more than 50\% of its mass (Guillochon \& Ramirez-Ruiz 2013).}$, 
respectively (see Figure 1 of Tchekhovskoy et al. 2014; see also the gray data points in the
top panel of Figure 1 of this paper). The X-ray intensity of the source drops abruptly 
by a factor of $\sim 170$ over a timescale of $\Delta t/t \la 0.2$ roughly a year after the 
disruption (see Figure 4 of Zauderer et al. 2013). This has been attributed to jet turnoff 
when the mass accretion rate dropped below the Eddington value, 
$\dot{M}_{\rm Edd} = L_{\rm Edd}/\eta c^{2}$, where $L_{\rm Edd}$, $\eta$, and $c$ are the Eddington 
luminosity, radiative efficiency, and speed of light, respectively (Zauderer et al. 2013).

Radio emission was detected from the source $\sim 0.9$\,d (rest frame) after 
its discovery in the hard X-rays (Zauderer et al. 2011). This early stage radio emission has 
been argued to represent relativistic jetted emission directly pointed along our line of sight 
(Zauderer et al. 2011). A follow-up radio campaign showed that the radio emission brightened 
starting about one month after discovery (observer frame; Berger et al. 2012). Berger et 
al. (2012) interpret this increase in energy as slower ejecta catching up with the forward 
shock at late times, although other explanations also exist (e.g., Barniol Duran \& Piran 2013).

Sw~J1644+57 is an exceptional TDE with signatures of a strong jet. Unfortunately, its host 
galaxy has a large line-of-sight extinction (Levan et al. 2011), making it challenging to study the 
evolution of the accretion disk expected to be observable in the ultraviolet/optical/infrared (UVOIR; e.g., Strubbe 
\& Quataert 2009; Lodato \& Rossi 2011). 

Although we have learned a great deal from Sw~J1644+57, the question of what aspect 
makes it conducive to produce a relativistic jet still remains. To answer this question, 
one approach would be to build a census of Sw~J1644+57-like sources. It has also been suggested 
that Sw~J1644+57 could represent a tidal disruption of a white dwarf by a member of the 
long-sought intermediate-mass black holes (IMBHs; mass range of a few $\times 10^{2-5}$\,M$_{\odot}$; 
Krolik \& Piran 2011). With only a handful of strong cases of such black holes known thus far 
(e.g., ESO HLX X-1: Farrell et al. 2009; Webb et al. 2012; M82 X-1: Kaaret et al. 2009; Pasham, 
Strohmayer, \& Mushotzky 2014), studying such systems could provide insight into weighing and 
hence identifying such unique objects.

Soon after the 2011 March 25 discovery of Sw~J1644+57, {\it Swift} discovered another 
transient, {\it Swift} J2058.4+0516 (hereafter Sw~J2058+05), on 2011 May 17 (Cenko et al. 
2012; hereafter C12). An early-time ($\la 2$ months since discovery), multiwavelength 
study showed a number of similarities with Sw~J1644+57 (C12). More specifically, Sw~J2058+05 
occupied the same location in the X-ray versus optical luminosity plot as Sw~J1644+57, and 
its early-phase (20\,d after outburst; rest frame) radio, UVOIR, and X-ray spectral 
energy distribution (SED) was similar to that of Sw~J1644+57 (see Figures 4 \& 5 of C12). More 
importantly, strong radio emission coincident with the X-rays was detected $\sim 20$\,d 
after the initial trigger, suggesting relativistic ejecta (C12). Unlike Sw~J1644+57, 
Sw~J2058+05 shows no evidence for line-of-sight extinction (C12), so we can study the system 
at UVOIR wavelengths in more detail. If it can be established that 
Sw~J2058+05 behaves analogously to Sw~J1644+57, then we can start to gain confidence that 
there is a class of such relativistic TDEs. This paper is a follow-up work to C12 and we 
address the remaining questions. (1) How does Sw~J2058+05 evolve on longer timescales? (2) 
Assuming the UVOIR can be modeled with a single-temperature blackbody, how do the 
properties of the putative blackbody evolve on these timescales? (3) Is the emission 
consistent with originating from the center of the host galaxy? (4) What is the mass of 
the disrupting black hole?

The paper is arranged as follows. In \S 2, we discuss the details of our X-ray, UVOIR, 
and radio observations. The results and the analysis are described in \S 3, while we discuss the 
similarity between Sw~J2058+05 and Sw~J1644+57, estimate the black hole mass,
and so on in \S 4. We give the main conclusions of this study in \S 5. Throughout this paper,
we adopt a standard $\Lambda$CDM cosmology with $H_{0}$ = 71 km s$^{-1}$ Mpc$^{-1}$, $\Omega_{m}$ 
= 0.27 and $\Omega_{\Lambda}$ = 1 - $\Omega_{m}$ = 0.73 (Spergel et al. 2007).

\section{Data Primer}
\subsection{X-ray Data}
The X-ray data of Sw~J2058+05 used in this study were acquired with three different instruments:  
 the X-Ray Telescope (XRT; Burrows et al. 2005) on the {\it Swift} Gamma-Ray Burst 
Explorer (Gehrels et al. 2004), the European Photon Imaging Camera (EPIC; Str{\"u}der et al. 
2001; Turner et al. 2001) on the {\it XMM-Newton} Observatory (Jansen et al. 2001), and 
the Advanced CCD Imaging Spectrometer (ACIS; Garmire et al. 2003) on the {\it Chandra} X-ray 
observatory (Weisskopf et al. 2002). We describe the data from each of these facilities below.

\subsubsection{{\it Swift}/XRT Observations}
Sw~J2058+05 was discovered by the BAT (Barthelmy et al. 2005) onboard {\it Swift} on 2011 May 
17 (Krimm et al. 2011). Soon after the BAT detection, starting 2011 May 27, a Target of Opportunity 
(ToO) program was initiated to monitor this source. Between 2011 May 27 and 2012 July 19 (a 
temporal baseline of 419\,d), {\it Swift} observed the source on 32 occasions for about 2--3\,ks per observation. 
The data from the first 60\,d of this monitoring program have already been reported by C12. 
They find that the source's hard X-ray flux falls below the BAT detection limits soon after 
reaching its peak luminosity (see the top panel of Figure 1 of C12). Here we extended the 
analysis to late times and use only the XRT X-ray (0.3--10\,keV) data from {\it Swift}.

We started our data analysis with the raw, level-1 XRT data products. Using the latest HEASARC 
calibration database (CALDB version 20140709) files, we ran {\it xrtpipeline} to extract the 
level-2 (scientific) event files. As suggested by the XRT data-analysis guide\footnote{
See \url{http://www.swift.ac.uk/analysis/xrt/lccorr.php}.}, we extracted the exposure maps to take into 
account the bad pixels and columns ({\it xrtpipeline} with the qualifier {\it createexpomap=yes}). 
These exposure maps were then used to correct the ancillary response files (arfs: effective area, 
using {\it xrtmkarf}) of each of the 32 observations.

Twenty two of the monitoring observations were taken in the photon-counting (PC) mode, with the remainder 
in the windowed-timing (WT) mode (see Table 1 for more details). As recommended by the XRT user 
guide, we only used events with grades 0--12 in the case of PC-mode observations and grades 
0--2 for WT-mode data. We then used {\it XSELECT} to extract energy spectra from each of the 
individual observations. For the PC-mode data we extracted the source spectra from a circular 
region centered on J2000.0 coordinates $\alpha = 20^{\rm h}58^{\rm m}19.90^{\rm s}$ and $\delta = 
+05^{\circ}13'32\farcs0$, as derived by C12 using the {\it Chandra}/High-Resolution Camera (HRC) 
data. We chose an extraction radius of 47\farcs1 to include roughly 90\% (at 1.5\,keV) of the 
light from the source (as estimated from XRT's fractional encircled energy function). A background 
region free of point sources was extracted from a nearby area. Given the low count rates, we 
chose a radius twice that of the source region in order to better estimate the background. For 
the WT-mode source region we chose a square box of width 94\farcs3 and oriented along the roll 
angle of the spacecraft --- that is, parallel to the WT-mode readout streak. Background was estimated 
from two square regions (width = 94\farcs3) on either side of the source region. The orientation 
of the square regions (both the source and the background) was adjusted between individual 
observations to align with the roll angle of the spacecraft during that particular exposure.

\subsubsection{{\it XMM-Newton} Observations}
{\it XMM-Newton} observed Sw~J2058+05 on three occasions (177, 179 and 340 d after the BAT detection; 
see Table 1 for further details). For the current study, 
we only used data acquired by EPIC, and both the ``pn'' and MOS data were used to 
achieve a higher signal-to-noise ratio (SNR). We started our analysis with the raw observation 
data files (ODF) and reprocessed them using {\it XMM-Newton}'s Standard Analysis System's (SAS) 
tools {\it epproc} and {\it emproc} for the pn and the MOS data, respectively. The standard data 
filters of {\it (PATTERN$\leq$4)} and {\it (PATTERN$\leq$12)} were used for the pn and the MOS 
data, respectively, and we only considered events in the energy range 0.3--10.0\,keV. All the 
time intervals of prominent background flaring were excluded from the analysis. The source events 
were extracted from a circular region of radius $33''$. This choice was made to include roughly 
90\% of the light from the source as estimated from the fractional encircled energy of the EPIC 
instruments. A background region of similar radius was chosen from a nearby uncontaminated region.

\subsubsection{{\it Chandra}/ACIS Observations}
{\it Chandra} observed Sw~J2058+05 on four occasions. One of the observations was during the 
early phase of the outburst (C12), while the others were carried out on days 685, 
896, and 899 after the initial BAT detection. Since we are interested in the late-time properties 
of the source, we only utilized the last three observations taken with ACIS. More details 
about these observations can be found in Table 1. Similar to the XRT and the EPIC data, we 
started our analysis with level-1 (secondary) data and reprocessed them using {\it Chandra}'s 
data-analysis system (CIAO) task {\it chandra\_repro} to account for any calibration changes 
that may have occurred since the epochs of these observations. Standard data filters were used 
for reprocessing. All further analysis was carried out on these level-2 event files.

\subsection{Ground-Based Optical Photometry Data}
Soon after discovery, we started a campaign to carry out multiband photometry of 
Sw~J2058+05 in the UV, optical, and NIR wavebands using 
multiple instruments. These include the High Acuity Wide field K-band-Imager (HAWK-I; Pirard 
et al. 2004) and the FOcal Reducer and Spectrograph 2 (FORS2; Appenzeller et al. 1998) on the 
8.2\,m Very Large Telescope (VLT), and the Gemini Multi-Object Spectrograph (GMOS; Hook et al. 
2004) mounted on the 8\,m Gemini-South telescope. VLT data were reduced via the standard 
instrument pipelines for FORS and HAWK-I in {\it esorex}, while Gemini data were
processed using the {\it gemini IRAF}\footnote{IRAF is distributed by the National 
Optical Astronomy Observatory, which is operated by the Association of Universities 
for Research in Astronomy (AURA) under a cooperative agreement with the National 
Science Foundation (NSF).} package.  Photometric calibration was
performed relative to nearby point sources from SDSS (optical) and 2MASS (NIR). The resulting 
photometry, all in the AB magnitude system, is presented in Table 2.  The reported magnitudes
are not corrected for 
foreground Galactic extinction along the line of sight to Sw~J2058+05 [$E(B-V) = 0.095$  
mag; Schlafly \& Finkbeiner 2011], but such corrections were applied before all
subsequent analysis. The observations prior to 2011 August 12 can be found in Table 1 
of C12.

\subsection{HST Observations}
We observed the location of Sw~J2058+05 with the Wide Field Camera 3 (WFC3) on the {\it Hubble 
Space Telescope} (\textit{HST}) in three separate epochs: 2011 Aug. 30, 2011 Nov. 30 (Proposal 
GO-12686; PI Cenko) and 2013 Dec. 10 (Proposal GO-13479; PI Levan). Observations were obtained with 
the $F160W$ filter through the IR channel in all three epochs, as well as with the $F475W$ 
filter through the UVIS channel in the first two epochs. An additional epoch of imaging was 
obtained on 2014 Aug. 31 in the $F606W$ filter with the Wide Field Camera (WFC) detector on the 
Advanced Camera for Surveys (ACS).  These data were downloaded after on-the-fly processing
from the \textit{HST} archive, and subsequently drizzled using \textit{astrodrizzle} 
(Fruchter \& Hook 2002) to final pixel scales of 65\,mas ($F160W$), 30\,mas ($F475W$),
and 33\,mas ($F606W$).  We performed aperture photometry at the location of 
Sw~J2058+05 in all images using the prescriptions from the various \textit{HST} handbooks. The 
resulting photometry, all corrected to the AB system, is displayed in Table 2.

\subsection{Optical and Near-Infrared Spectra}
We obtained optical and NIR spectra of Sw~J2058+05 with the Low-Resolution Imaging 
Spectrometer (LRIS; Oke et al. 1995) on the 10\,m Keck I telescope, FORS2 on the 8\,m VLT 
UT1 (Antu), and the XSHOOTER (Vernet et al. 2011) spectrograph on the 8\,m 
VLT UT2. Details of the configuration for each spectrum are provided in Table 3.  For
the Keck/LRIS and FORS2 data, one-dimensional spectra were optimally extracted, a wavelength 
solution was generated from observations of lamps, and flux calibration was performed 
via spectrophotometric standards.  The XSHOOTER spectra were processed through the
\textit{reflex} environment.  For all spectra the slit was oriented at the parallactic 
angle to minimize losses caused by atmospheric dispersion (Filippenko 1982).

\subsection{Radio Data}
We obtained a single epoch of imaging of Sw~J2058+05 with the National Radio
Astronomy Observatory's (NRAO\footnote{The National Radio Astronomy Observatory is a 
facility of the NSF operated under cooperative agreement by 
Associated Universities, Inc.}) Very Long Baseline Array (VLBA) to search for spatially 
extended radio emission (project code BC0199). Observations were obtained on 
2011 Aug. 12 ($\Delta t \approx 40$\,d after discovery, in the rest frame) at central 
frequencies of 8.4 and 22\,GHz with a recording rate of 512\,Mb\,s$^{-1}$.  All 10 
stations (SC, HN, NL, FD, LA, PT, KP, OV, BR, and MK) were planned for both frequencies; 
however, the NL station was lost for our 8\,GHz observation (owing to a receiver
problem).

Initial data processing was performed using the AIPS software package
(Greisen 2003).  J2101+0341 was used for primary phase and astrometric calibration,
while J2050+0407 and J2106+0231 were used as secondary calibrators and for 
evaluation and correction of tropospheric effects on astrometry.  The
resulting images achieved an angular resolution of $\sim 1$\,mas at
8.4\,GHz and 0.3\,mas at 22\,GHz.

A faint ($f_{\nu} = 350 \pm 70$\,$\mu$Jy), unresolved source is detected in the 
8.4\,GHz image at the J2000.0 position $\alpha = 20^{\mathrm{h}} 58^{\mathrm{m}}
19.897282^{\mathrm{s}}$ $\pm$ 0.000006$^{\mathrm{s}}$, $\delta = +05^{\circ} 13\arcmin 32\farcs24306$ 
$\pm$ 0.00016${\mathrm{''}}$\footnote{The reported uncertainties in both RA and DEC are the statistical 
errors obtained from fits to VLBA data}. No emission
is detected at this location in the 22\,GHz image to a 3$\sigma$ upper limit of 
$f_{\nu} < 580$\,$\mu$Jy.  Both measurements suggest a decline in radio luminosity
by a factor of a few from VLA observations of Sw~J2058+05 presented by C12 ($\sim 20$\,d rest frame).

\section{Analysis}

This section is divided into five parts: (1) we show the long-term X-ray light curve of 
Sw~J2058+05 and compare its behavior with that of Sw~J1644+57; (2) we carry out astrometry 
using {\it HST} and VLBA to pin down Sw~J2058+05's location within its host galaxy; 
(3) we study the evolution of the UVOIR SED;
(4) we analyze the late-time optical spectra; and (5) we consider limits 
on the size of the radio-emitting region.

\subsection{Long-term and Short-term X-ray Light Curves}
The individual {\it Swift}/XRT observations do not have enough counts to constrain 
the source's spectral parameters. Hence, we extracted an average energy spectrum 
by combining all of the XRT PC-mode data\footnote{We excluded the WT-mode data to avoid 
any systematics caused by the low-energy spectral residuals as described in {\it Swift} 
XRT digest at \url{http://www.swift.ac.uk/analysis/xrt/digest\_cal.php}.}. This was 
achieved by first extracting a source spectrum and a background energy spectrum from each of 
the 22 observations (most of these observations were at epochs 25--86\,d 
after discovery; rest frame) and then combining them all using the ftool {\it sumpha}. 
Similarly, we combined all of the individual ancillary response files, weighted by total 
counts per observation, using the ftool {\it addarf}. The response files (RMF) were 
averaged using the ftool {\it addrmf}. The combined spectrum was then rebinned using the {\it grppha} tool to 
have a minimum of 25 counts per spectral bin. 

With the 
latest version of the X-ray spectral fitting package {\it XSPECv12.8.2} (Arnaud 1996), we 
then fitted this combined 0.3--10\,keV energy spectrum with a power-law model modified by 
absorption ({\it phabs$*$zwabs$*$pow} in XSPEC). The Galactic column density was fixed 
at $0.088 \times 10^{22}$\,cm$^{-2}$ (Kalberla et al. 2005; Willingale et al. 
2013)\footnote{See \url{http://www.swift.ac.uk/analysis/nhtot/index.php}.}, while the 
power-law index and the intrinsic absorption column at $z = 1.1853$ (C12) 
were free to vary. The best-fit power-law index and intrinsic absorption column 
density were $\Gamma = 1.47 \pm 0.08$ and $n_{\rm H} = (0.30 \pm 0.15) \times 10^{22}$\,atoms\,cm$^{-2}$, respectively (with a reduced $\chi^2 = 0.74$ for 102 degrees of freedom).
 
We then used these best-fit power-law model parameters (fixing the power-law index and the absorbing 
column density but keeping the power-law normalization free) and extracted the source 
flux from each of the individual observations. We only considered observations with a 
total number of counts greater than 50. In cases where the total number of counts was 
less than 50, we averaged neighboring observations. The best-fit absorbed power-law model 
(with fixed power-law index and absorbing column) yielded a reduced $\chi^2$ in an 
acceptable range of 0.5--1.3 for these individual epochs. 

We then fitted each of the three {\it XMM-Newton}/EPIC (both pn and MOS simultaneously) 
X-ray spectra of Sw~J2058+05 using the same model as above ({\it phabs$*$zwabs$*$pow}). We 
generated the EPIC response files using the $arfgen$ and $rmfgen$ tools which are part 
of {\it XMM-Newton}'s SAS software. Given that each of these observations had total counts 
in excess of 1600, we left all the model parameters free to vary except for the redshift 
and the Milky Way column density. The best-fit model parameters are indicated in Table 4. 
It is interesting to note that while the best-fit absorbing column densities are consistent 
with the value derived from the combined {\it Swift} XRT data acquired at early times, the 
power-law indices are slightly steeper at late times. The luminosity values derived from 
modeling the {\it XMM-Newton} spectra are indicated by the magenta 
squares in Figure 1. 

The source was not detectable in the {\it Chandra/ACIS} images with the naked eye. Nevertheless, 
using the CIAO task {\it srcflux}, we estimated an upper limit to the 0.3--10\,keV X-ray flux
for Poisson statistics. 
In doing so, we assumed that the source spectrum is defined by an absorbed power-law model 
with the parameters estimated from the {\it XMM-Newton} data (see Table 4). The power-law index and the
intrinsic absorption column density were set to 1.79 and $0.19 \times 10^{22}$\,atoms\,cm$^{-2}$, 
respectively (mean of the {\it XMM-Newton} values). The isotropic luminosity upper limits are 
indicated by the blue squares in Figure 1.

In addition, we studied the short-term variability of the source on  timescales of 
a few hundred to a few thousand seconds using the {\it XMM-Newton} data. We first extracted a 
combined EPIC (pn and MOS) light curve from each of the three {\it XMM-Newton} observations. 
One such light curve (black) along with the background (red) binned with a time resolution 
of 500 s is shown in Figure 2. It is clear that the source varies significantly, with 
the most drastic variation around 32,000\,s when the overall count rate changes by a 
factor of 2.5 within a timescale of $\la 1000$\,s. To further confirm the variability,  
we modeled the light curve with a constant. The best-fit model gave 
0.073 counts s$^{-1}$ with a reduced $\chi^2$ of 2.3 ($\chi^2 = 236$ for 102 
degrees of freedom). Again, this suggests that a constant count rate model is strongly 
disfavored.  Rapid X-ray variability on similar timescales has also been observed from 
Sw~J1644+57 (e.g., Krolik \& Piran 2011) and also nonrelativistic TDE candidates such as
SDSS~J120136.02+300305.5 (see Figure 5 of Saxton et al. 2012).

Finally, to test for any possible coherent oscillations in the X-rays (0.3--10\,keV), we extracted a 
power-density spectrum using the longest {\it XMM-Newton} observation (ObsID: 0694830201) with an
effective exposure of roughly 48\,ks. We find that the power spectrum is flat (white noise) 
and is consistent with being featureless (see Figure 3).

\subsection{HST Astrometry}
Dynamical friction within a galaxy ensures that supermassive black holes sink to the center 
within a few Gyr after formation (e.g., Equation 4 of Miller \& Colbert 2004). 
Therefore, if Sw~J2058+05 is an event caused by a supermassive black hole, it should arise from 
the center of the host galaxy.  To constrain the (projected) offset between the 
transient emission from Sw~J2058+05 and its underlying host, we took three approaches.

First, we compared the VLBA position for Sw~J2058+05 (\S2.5) with the host localization
derived from \textit{HST}.  We used the $F606W$ image from 2014 for this purpose (as opposed to
the $F160W$ images obtained in Dec. 2013), owing to its higher SNR and smaller native pixel 
scale.  While the VLBA position for Sw~J2058+05 is the most precise available, the dominant 
source of uncertainty results from alignment of the \textit{HST} images onto the FK5 reference 
grid using common point sources from 2MASS (60\,mas in each coordinate). After alignment, 
we measured a position for the host centroid in the \textit{HST} images of $\alpha = 20^{\mathrm{h}} 58^{\mathrm{m}}
19.898^{\mathrm{s}}$, $\delta = +05^{\circ} 13\arcmin 32\farcs30$.  As this is offset
from the VLBA position by 58\,mas, we conclude that the radio position is consistent
with the host nucleus, within our uncertainties.

Next, we performed digital image subtraction on our $F160W$ images 
obtained on 2011 Aug. 30 (top-left panel of Figure 4) and 2013 Dec. 10 (top-right panel of Figure 4) 
to more precisely constrain the \textit{relative} transient-host offset 
(e.g., Levan et al. 2011). The resulting subtraction image is displayed in the bottom-left panel  
of Figure 4. Assuming that the flux in the final epoch of imaging is dominated by the host 
galaxy, we measured a radial offset between the transient emission and the host centroid of 
0.34 pixels (i.e., 22\,mas). Including 
contributions to the relative astrometric uncertainty from image alignment (0.18 pixel in each 
coordinate) and measurement of the host centroid (0.10 pixel in each coordinate), we find a 
null probability of measuring such an offset of 27\% (assuming a Rayleigh distribution for 
the radial offset). Thus, we conclude that the transient emission is consistent with the host 
nucleus at this level of precision, as well. 

Finally, we measured the relative offset between the 2011 $F475W$ images of 
Sw~J2058+05 (dominated by transient emission) and our 2014 $F606W$ image of the field
(presumed to be host dominated).  We find that the centroids in the two images are offset
by 10\,mas, while our alignment uncertainty is only 15\,mas in each coordinate.
This method offers the most precise constraint on the relative transient-host offset,
and we formally place a 90\% confidence limit of $\Delta \theta \lesssim 45$\,mas,
corresponding to a projected distance of $\lesssim 400$\,pc at $z = 1.1853$.  
This is comparable to the limits on the transient-host offset derived for Sw~J1644+57 
($d < 150$\,pc (1$\sigma$) at $z = 0.354$; Levan et al. 2011).

\subsection{Temperature and Radius Evolution of the Blackbody}
The UVOIR SED of Sw~J2058+05 at early times
($\Delta t \lesssim 1$\,yr, observer frame) is quite blue, significantly more so
than one would expect from simple forward-shock models (e.g., Granot \& Sari 2002).  
Motivated by the observed SEDs in nonrelativistic TDEs (e.g., Gezari et al. 2012),
we fit, wherever possible, the UVOIR SEDs with a single-temperature blackbody.
The best-fit model parameters from the six epochs are indicated in Table 5. 
Including host-galaxy extinction as an additional free parameter 
in modeling these SEDs did not improve the fits. Formally, we limit the host extinction to 
$A_V \la 0.2$ mag (90\% confidence), assuming it has an extinction law similar to 
that in the Milky Way (Pei 1992).  

All of the SEDs along with the best-fit model are shown in the top panel of Figure 5. 
We show the evolution of the temperature and the radius of the blackbody in the bottom-left 
and bottom-right panels, respectively. There is clear evidence for a decrease in the 
blackbody temperature at late times before the X-ray flux drops off, and marginal evidence
for an increase in the radius. But given the large error bars in the radii, we cannot strongly
rule-out the possibility that the radius remains constant throughout.  We also note, however, that with
reduced $\chi^{2}$ values as low as we find in several epochs, the quoted uncertainties
should be treated with some degree of caution.  Regardless, it is clear that 
the emission has become much redder in our final epoch, with a largely flat SED 
in $\nu L_{\nu}$.

\subsection{Optical/Near-Infrared Spectra}
Our highest-SNR spectrum, obtained with Keck/LRIS on 2011 Aug. 28, is plotted in Figure 6. 
We also fit our Keck/LRIS spectra to single-temperature blackbody models, and find
$T_{\mathrm{BB}} = (1.8 \pm 0.2) \times 10^{4}$\,K on 2011 Aug. 2 and 
$T_{\mathrm{BB}} = (2.3 \pm 0.1) \times 10^{4}$\,K on 2011 Aug. 28 (solid green line
in Figure 6).  These results are largely consistent with the values derived from 
our broadband photometry, providing additional confidence in the above analysis.

For comparison, in Figure 6 we also plot the composite quasar (QSO) spectrum from SDSS (Vanden 
Berk et al. 2001), and a spectrum taken near maximum light of the TDE PS1-10jh (Gezari et al. 
2012). In all cases the spectra of Sw~J2058+05 
are dominated by a blue, featureless continuum. No obvious emission or absorption features are 
detected in any spectra, with the exception of the initial spectrum from 2011 June 1 presented 
in C12, from which the redshift of $z$ = 1.1853 was derived from narrow \ion{Mg}{2} and \ion{Fe}{2} 
absorption lines. Clearly, the strong, broad emission lines
that dominate QSOs in the near-UV (e.g., \ion{Mg}{2}, \ion{C}{3}], and \ion{C}{4}) are not 
present in our spectra of Sw~J2058+05. In addition to a hot ($T_{\mathrm{BB}} \approx 3 
\times 10^{4}$\,K) blackbody continuum, PS1-10jh displayed high-ionization \ion{He}{2} 
$\lambda$4686 and $\lambda$3203 emission lines. Our Keck/LRIS spectra do not probe
sufficiently far into the rest-frame optical to cover the stronger \ion{He}{2} $\lambda$4686 feature. 
We see no evidence for broad emission at this location in our XSHOOTER spectra; however, the 
SNR is quite low in these data. A number of optically discovered TDEs also display broad 
H$\alpha$ emission (Arcavi et al. 2014; Holoien et al. 2014), although it is unclear if the
presence/absence of H is due to properties of the disrupted star (Gezari et al. 2012) or the 
radial extent of the newly formed accretion disk (Guillochon et al. 2014). Again, we detect 
no evidence for broad emission lines at rest-frame H$\alpha$ (or any other Balmer lines, for 
that matter), but are limited by the low SNR at these wavelengths.

We can also limit the presence of narrow, nebular emission 
lines from the underlying host galaxy. In particular, we do not detect either [\ion{O}{2}] 
$\lambda$3727 or H$\alpha$. If we assume unresolved emission lines at these wavelengths, we 
calculate limiting flux values of $f$(\ion{O}{2}) $< 4.8 \times 10^{-17}$\,erg\,s$^{-1}$\,cm$^{-2}$  
and $f$(H$\alpha$) $< 6.5 \times 10^{-17}$\,erg\,s$^{-1}$\,cm$^{-2}$. Using the relations from 
Kennicutt (1998) between emission-line luminosity and star-formation rate, we limit
the presence of recent star formation in the host of Sw~J2058+05 to be $\lesssim 5$\,M$_{\odot}$\,yr$^{-1}$ 
(uncorrected for extinction).  This is consistent with an estimate of the 
star-formation rate derived from the UV ($F606W$) luminosity of the host galaxy, for 
which we find 0.8\,M$_{\odot}$\,yr$^{-1}$ (using the calibration from Kennicutt 1998).

\subsection{Size of the Radio-Emitting Region}
The detection of radio emission from Sw~J2058+05 confirms the presence of nonthermal
electrons in the circumnuclear ejecta.  We can apply standard equipartition arguments
(Readhead 1994; Kulkarni et al. 1998) to place a lower limit on the size of the
radio-emitting region.  Using the formulation valid for relativistic outflows from
Barniol Duran et al. (2013), our VLBA detection at $\Delta t \approx 40$\,d (rest frame)
implies $R_{\mathrm{eq}} \gtrsim 7 \times 10^{16}$\,cm.  Similarly, these observations,
though not as constraining as those presented by C12\footnote{Applying the same 
formulation to the VLA data from C12, we find $R_{\mathrm{eq}} \gtrsim 6 \times 
10^{16}$\,cm, $\Gamma_{\mathrm{eq}} \gtrsim 1.5$, and $E_{\mathrm{eq}} \gtrsim 
4 \times 10^{49}$\,erg.}, imply at least transrelativistic expansion ($\Gamma_{\mathrm{eq}} 
\gtrsim 0.6$) from an energetic outflow ($E_{\mathrm{eq}} \gtrsim 3 \times 
10^{49}$\,erg).

The above limit on the physical size of the radio-emitting region corresponds to 
a lower limit on the angular size of $\Theta \gtrsim 3 \Psi$\,$\mu$as, where $\Psi$ is 
the jet opening angle.  For any feasible jet opening angle, this result is consistent
with the unresolved nature of the source in the VLBA imaging ($\Theta \lesssim 1$\,mas).

\section{Discussion}

\subsection{Radiation Mechanisms and the Broadband SED}
To better understand the nature of Sw~J2058+05, we first consider the origin of the
 emission in the three regimes probed here: radio, UVOIR, and X-ray.  We
derived a robust lower limit on the size of the radio-emitting region (based solely
on equipartition arguments in \S3.5), $R_{\mathrm{radio}} \gtrsim 7 \times 10^{16}$\,cm.
Together with more stringent limits on the bulk Lorentz factor from C12
($\Gamma \gtrsim 1.5$), we conclude that the radio emission is generated by the 
forward shock of a newly formed, at least mildly relativistic jet.  An identical
conclusion was reached by several authors (e.g., Zauderer et al. 2011; Bloom et al.
2011) in the case of Sw~J1644+57.

The X-rays, on the other hand, must clearly have a distinct origin.  The rapid 
variability on a rest-frame timescale of $\lesssim 500$\,s require the size of the
X-ray-emitting region to be $R_{\mathrm{X-ray}} \lesssim c\, \delta t \approx 2 \times
10^{13}$\,cm.  This clearly rules out a forward-shock origin.  However, the tremendous
peak X-ray luminosity, many orders of magnitude above Eddington for any feasible
black hole, suggests some association with the newly formed jet (as does the 
rapid turnoff; see below).  One possibility is that the X-rays are generated in the
base of the jet (e.g., Bloom et al. 2011; Zauderer et al. 2011), though the process
by which this occurs remains a mystery.  Again, the analogy with Sw~J1644+57 holds well.

Finally, we have demonstrated that the UVOIR data, both photometry and
spectra, are well fit by a single-temperature blackbody with $T_{\mathrm{BB}} 
\approx$ few $\times 10^{4}$\,K.  The inferred blackbody radius, which appears to remain
roughly constant, is $R_{\mathrm{opt}} \approx 10^{15}$\,cm.  Together with the 
long-lived blue colors, the radius also seems to disfavor a forward-shock origin
for the UVOIR component.  Similarly, the derived blackbody spectrum severely
underpredicts the observed X-ray flux.

Instead, these values are consistent with spectral studies of nonrelativistic 
TDE candidates in the literature with apparent blackbody temperatures and radii in the range 
of (1--10) $\times 10^{4}$\,K and (0.1--20) $\times 10^{15}$\,cm, respectively (e.g., Gezari et 
al. 2009b, 2012; Armijo \& de Freitas Pacheco 2013; Guillochon et al. 2014; Chornock et 
al. 2014; Cenko et al. 2012b; Holoien et al. 2014; Arcavi et al. 2014), although there are some TDE candidates 
that tend to show higher disk temperatures of $\ga 10^{5}$\,K accompanied by smaller emitting 
regions of size $\la 10^{13}$\,cm (e.g., Gezari et al. 2008).  However, for any 
plausible black hole mass, the blackbody radius is orders of magnitude larger than
the radius at which disruption should occur.  Such large radii have been attributed
to reprocessing in some external region (see, for example, the numerical simulations
of Guillochon et al. 2014).

It is important to note here, that while Sw~J1644+57 lacked detectable UV and optical emission,
the high degree of polarization observed in the NIR was attributed to jetted emission
from the forward shock (Wiersema et al. 2012), and not from the (presumably largely 
isotropic) accretion disk.  Naively, unless the reprocessing region was nonisotropic,
we would expect a low degree of optical polarization from Sw~J2058+05 if this
simplistic picture is correct.  For future relativistic TDE candidates, polarization 
observations would be an important test of this model.

\subsection{Energetics}
Using the best-fit blackbody luminosities and integrating the resulting light curve (using 
the trapezoidal rule) in the rest frame between epochs 5.7 and 181.4\,d, we estimate the 
total UVOIR energy radiated to be $\sim 5 \times 10^{51}$\,erg. Similarly, 
we integrated the X-ray light curve (top panel of Figure 1) and estimate the total isotropic 
energy to be $\sim 4 \times 10^{53}$\,erg. Assuming an opening angle of $\sim 0.1$ rad, 
similar to what has been estimated for Sw~J1644+57 (Zauderer et al. 2013; Metzger et al. 2012), 
we measure the total, beaming-corrected X-ray energy output to be $\sim 4 \times 10^{51}$\,erg. 
The bolometric luminosity is, however, expected to be a factor of a few higher than the 
X-ray luminosity. Assuming the bolometric value is a factor of 3 (similar to that of Sw~J1644+57; 
Burrows et al. 2011), one can estimate the total accreted mass onto the black hole using 
Equation 5 of the supplemental information of Burrows et al. (2011). We find this value to be 
$\sim 0.1$\,M$_{\odot}$, which is comparable to Sw~J1644+57's 0.2\,M$_{\odot}$ (Burrows et al. 
2011; Zauderer et al. 2013), both appropriate for disruption of a $\sim 1$\,M$_{\odot}$ star. 

\subsection{Nature of the Rapid X-ray Dropoff}
The X-ray emission from Sw~J2058+05 drops abruptly between days 200 and 300 (rest frame), 
consistent with what was seen for Sw~1644+57 (top panel of Figure 1). More specifically, Sw~J2058+05's 
intensity decreases by a factor of $\ga 160$ within a span of $\Delta t/t \le 0.95$ compared to 
Sw~J1644+57's factor of $\sim 170$ decline over a span of $\Delta t/t \la 0.2$ (Levan \& Tanvir 
2012; Sbarufatti et al. 2012; Zauderer et al. 2013). Interestingly, in both of these 
sources, the X-ray dimming occurs on a comparable timescale after disruption. 

In the case 
of Sw~J1644+57, Zauderer et al. (2013) interpreted this sudden decrease in the flux as an 
accretion-mode transition from a super-Eddington to a sub-Eddington state. This is consistent 
with the transitioning timescale predicted from numerical simulations (e.g., Figures 
4 \& 2 of Evans \& Kochanek 1989 and De Colle et al. 2012, respectively). Assuming the 
same process is responsible for the abrupt flux change in Sw~J2058+05, we can attempt to 
estimate the mass of the black hole by equating the luminosity at turnoff to the 
Eddington luminosity. From the 
X-ray light curve (see Tables 1 \& 4), it is evident that the isotropic X-ray luminosity drops 
from $1.3 \times 10^{45}$\,erg\,s$^{-1}$ to less than $8.4 \times 10^{42}$\,erg\,s$^{-1}$, suggesting 
an Eddington value somewhere in between these two limits. Using these two values and assuming 
radiative efficiency, beaming angle, and bolometric correction values of 0.1, 0.1 rad,
and 30\%, respectively (similar to Sw~J1644+57; Burrows et al. 2011), we constrained the black 
hole mass $M_{\rm BH}$ to be $10^{4}$\,M$_{\odot} \la M_{\rm BH} \la 2 \times 10^{6}$\,M$_{\odot}$. 
Furthermore, numerical simulations suggest that the time to dropoff (transition from super-Eddington 
to sub-Eddington) since the disruption is shorter for more massive black holes (see Figure 2 of De 
Colle et al. 2012). The X-ray dropoff in Sw~J2058+05 occurs $\sim 100$\,d earlier than that in Sw~J1644+57, 
suggesting that its black hole may be more massive.

However, it is interesting to note that even the optical light curves of Sw~J2058+05 undergo  
an abrupt change during an epoch roughly consistent with the X-ray dimming (see the bottom panel of 
Figure 1). We find that the optical flux, for instance in the $r$ band, drops by a factor of at least 5 
within a narrow span of $\Delta t /t \approx 0.16$. We speculate, within the context of the 
following simple model, that the X-rays are coming from the base of the jet and the optical 
originates from the reprocessed UV/soft-X-ray disk photons in the ambient medium. In such a scenario, 
the proposed super-Eddington to sub-Eddington accretion transition would presumably change the 
accretion-disk structure to lower its emission, thus explaining the reduction in the amount of 
the reprocessed light. Obviously, the true situation is more complicated, with specific details about 
the radiative efficiency, beaming, and other factors. It can be better understood with more 
detailed modeling, but this is beyond the scope of the current paper.

On the other hand, the longterm X-ray light curve of Sw~J2058+05 does not exhibit the numerous sudden 
dips observed in Sw~J1644+57 (see Figure 1). In the case of Sw~J1644+57, it has been 
argued that the X-ray dips originate from jet precession and nutation, which causes it to briefly 
to go out of our line of sight (e.g., Saxton et al. 2012). We speculate, based on the lack of such 
dips in Sw~J2058+05, that its jet may be more stable compared to Sw~J1644+57. However given the poor 
sampling of the X-ray light curve, the current data cannot completely rule-out the presence of dips 
in Sw~J2058+05.

\subsection{Other $M_{\mathrm{BH}}$ Estimates}
The mass limits derived above based on the X-ray turnoff are consistent with 
other methods of estimating $M_{\mathrm{BH}}$.  First, we can use the X-ray variability
timescale to place an upper limit on black hole mass.  Equating the limit on the
size of the X-ray-emitting region with a Schwarzschild radius suggests a compact 
object of mass less than $5 \times 10^{7}$\,M$_{\odot}$.

Also, assuming that the optical flux in our final two {\it HST} epochs is dominated by the host
galaxy (and not transient emission), we can constrain the mass of the central  
supermassive black hole using the well-known bulge luminosity vs. black hole mass relations 
(e.g., Lauer et al. 2007). Neglecting for the moment K-corrections [aside from the 
cosmological $-2.5 \log(1+z)$ factor], the distance modulus at $z = 1.1853$ implies an 
absolute magnitude of $-18.7$ from the $F606W$ observation (approximately rest-frame $U$ band) 
and $-19.4$ from the $F160W$ observation (approximately rest-frame $I$ band). Both suggest $M_V 
\approx -19$\,mag, or an inferred supermassive black hole mass of $M_{\rm BH} \lesssim 3 \times 
10^{7}$\,M$_{\odot}$. While there is significant scatter in the bulge luminosity vs. black hole 
mass relation, our limits are conservative in the sense that they assume \textit{all} of the 
observed luminosity derives from the bulge (and none from, say, a disk). At the very least,  
we can robustly conclude that $M_{\rm BH} < 10^{8}$\,M$_{\odot}$, the limit above which a 
nonspinning black hole cannot tidally disrupt a solar mass main-sequence star (Rees 1988).

\section{Conclusions}
The goal of this work is to use multiwavelength data and study the long-term ($\sim 1$\,yr)
behavior of candidate relativistic TDE Sw~J2058+05. Our main conclusions are as follows.

(1) The long-term X-ray turnoff and the host-galaxy nuclear association of Sw~J2058+05 strengthen 
the similarity between Sw~J2058+05 and Sw~J1644+57.

(2) Rapid X-ray variability on a timescale $\la 500$\,s at late times (before the X-ray dropoff) 
suggests that X-ray photons originate near the black hole and not from a forward shock. 
If the X-rays were to come from the forward shock, they would vary on much longer timescales.

(3) Based on the blackbody modeling of the optical data of Sw~J2058+05 (in ways not possible with 
Sw~J1644+57 because of the large host-galaxy extinction), we find that the optical originates from farther 
out ($\sim 10^{15}$\,cm) than the X-rays. Also, the UVOIR SED modeling severely underpredicts the 
X-ray emission. Lastly, the early-time optical data did not show variability on timescales 
of a few 1000\,s, suggesting again an emission size of larger than a few 1000 light seconds. However, the 
X-rays originate very close to the black hole. We conclude based on these lines of evidence that the 
optical and the X-rays have distinct origins.

(4) The size of the optically emitting region of Sw~J2058+05 suggests that it originates from reprocessing.
The fact that reprocessing is seen in X-ray-selected events (as well as optical ones) suggests it is a
relatively common phenomenon. 

(5) In Sw~J2058+05-like events, the X-ray dropoff (both the flux or the timescale measurements) could 
be a probe of the black hole mass.

(6) These observations imply the need for improved modeling to better understand Sw~J2058+05-like 
events.

{\it Facilities:} \facility{Swift (XRT)}, \facility{XMM (EPIC)}, \facility{Chandra (ACIS)},
\facility{VLT (FORS2, HAWK-I, XSHOOTER)}, \facility{Keck (LRIS)}, \facility{Gemini:South
(GMOS-S)}, \facility{HST (WFC3, ACS)}, \facility{VLBA}


\acknowledgments
We thank the \textit{XMM-Newton} and \textit{HST} teams, in particular Project Scientist 
N.~Schartel and STScI director M.~Mountain, for the approval and prompt scheduling of
our DD requests.  We are also grateful to James Guillochon and Ryan Chornock for valuable
discussions. D.R.P. is grateful for valuable discussions with Sjoert van Velzen and Nick Stone. 
S.B.C. thanks the Aspen Center for Physics and NSF Grant \#1066293 for hospitality during 
the preparation of this manuscript. K.L.P. acknowledges support from the UK Space Agency.
Support for this work was provided by the National Aeronautics and Space Administration (NASA) 
through Chandra Award Number GO3-14107X issued by the Chandra X-ray Observatory Center, 
which is operated by the Smithsonian Astrophysical Observatory for and on behalf of NASA 
under contract NAS8-03060.  D.R.P. and S.B.C. also acknowledge support from \textit{HST} 
grant GO-13611-006-A. The work of A.V.F. was made possible by NSF grant AST-1211916, the 
TABASGO Foundation, and the Christopher R. Redlich Fund. A.V.F. and S.B.C. also acknowledge
the support of Gary and Cynthia Bengier. Finally, we would like to thank the referee for 
his/her careful comments and suggestions. 

The scientific results reported in this article are based in part on observations made by 
the {\it Chandra} X-ray Observatory, NASA/ESA {\it Hubble Space Telescope}, obtained from 
the Data Archive at the Space Telescope Science Institute, which is operated by the 
Association of Universities for Research in Astronomy, Inc., under NASA contract NAS 5-26555
and, {\it XMM-Newton}, an ESA science mission with instruments and contributions directly 
funded by ESA Member States and NASA. Some of the data presented herein were obtained at 
the W. M. Keck Observatory, which is operated as a scientific partnership among the California 
Institute of Technology, the University of California, and NASA; the observatory was made 
possible by the generous financial support of the W. M. Keck Foundation.  Also, observations 
were obtained at the Gemini Observatory, which is operated by the 
Association of Universities for Research in Astronomy, Inc., under a cooperative agreement 
with the NSF on behalf of the Gemini partnership: the National Science Foundation 
(United States), the National Research Council (Canada), CONICYT (Chile), the Australian 
Research Council (Australia), Minist\'{e}rio da Ci\^{e}ncia, Tecnologia e Inova\c{c}\~{a}o 
(Brazil) and Ministerio de Ciencia, Tecnolog\'{i}a e Innovaci\'{o}n Productiva (Argentina).
Also, based on observations made with ESO Telescopes at the La Silla or Paranal Observatories.
We acknowledge the use of public data from the Swift data archive.

\newpage
\begin{figure}

\begin{center}
\vspace{-.25in}
\includegraphics[width=5.5in, height=7.in, angle=0]{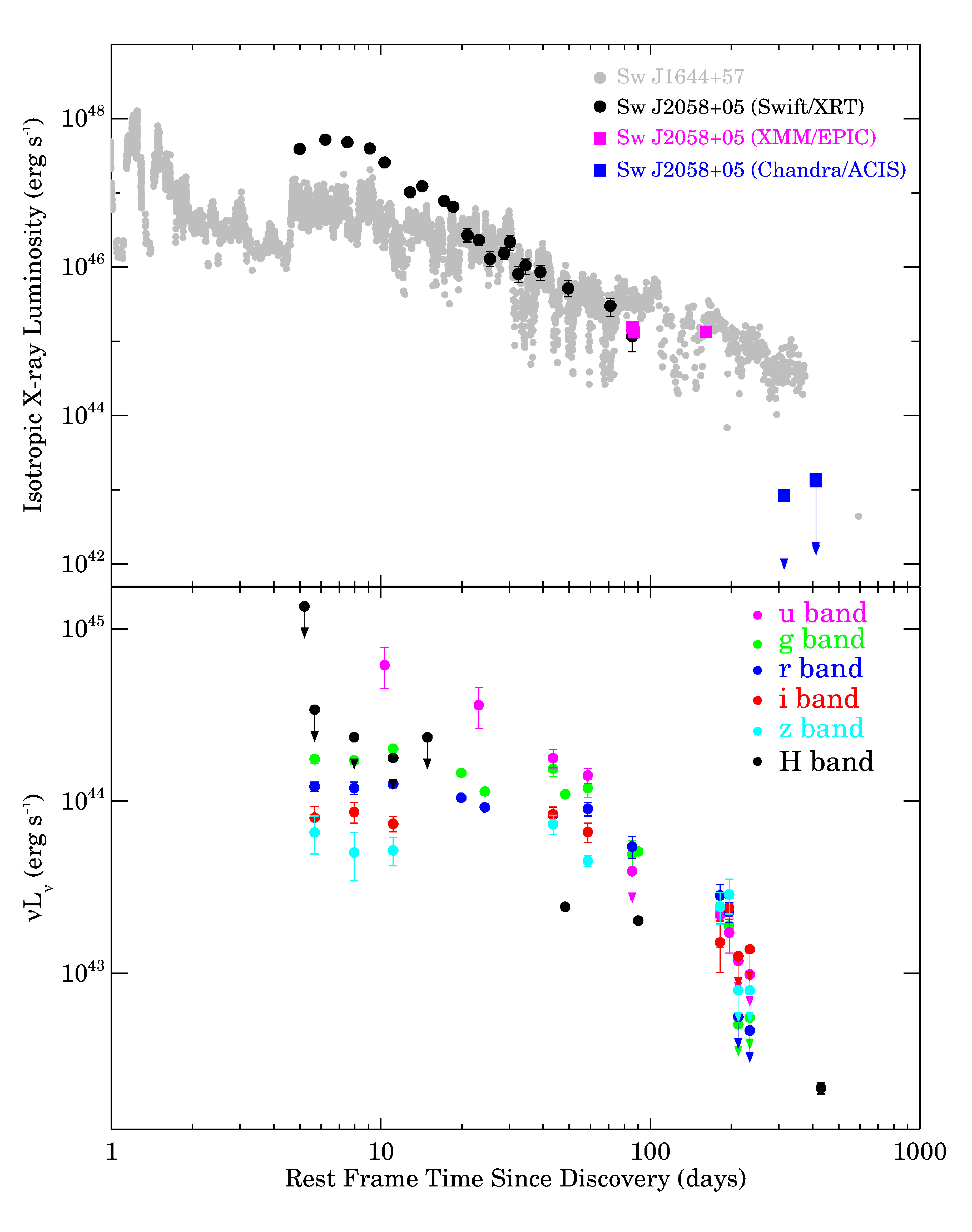}
\end{center}
{\textbf{Figure 1:} {\it Top panel:} Comparing the long-term X-ray (0.3--10\,keV) light curve of Sw~J2058+05 
(filled circles and squares) with that of Sw~J1644+57 (gray; adapted from Burrows et al. 
2011). An abrupt decline in the X-ray luminosity seen in Sw~1644+57 (Zauderer et al. 2013) 
is also evident in Sw~J2058+05. The magenta squares are 
flux estimates of Sw~J2058+05 from {\it XMM-Newton}/EPIC, while the blue are the upper limits 
from {\it Chandra}/ACIS (see Table 1). The discovery time of Sw~J2058+05 is not precisely 
constrained, but we refer to the time of discovery as 00:00:00 on 2011 May 17 (MJD = 55698) 
as per Cenko et al. (2012). The rest-frame time was thus estimated as (time $-$ 55698)/$(1+z)$, 
where $z = 1.1853$. {\it Bottom panel:} The long-term light curve of Sw~J2058+05 in various 
optical bands (data available in Table 2), showing a similar sharp decline as seen in the X-rays. 
}
\label{fig:figure1}
\end{figure}
\clearpage

\begin{figure}

\begin{center}
\includegraphics[width=6.5in, height=4.5in, angle=0]{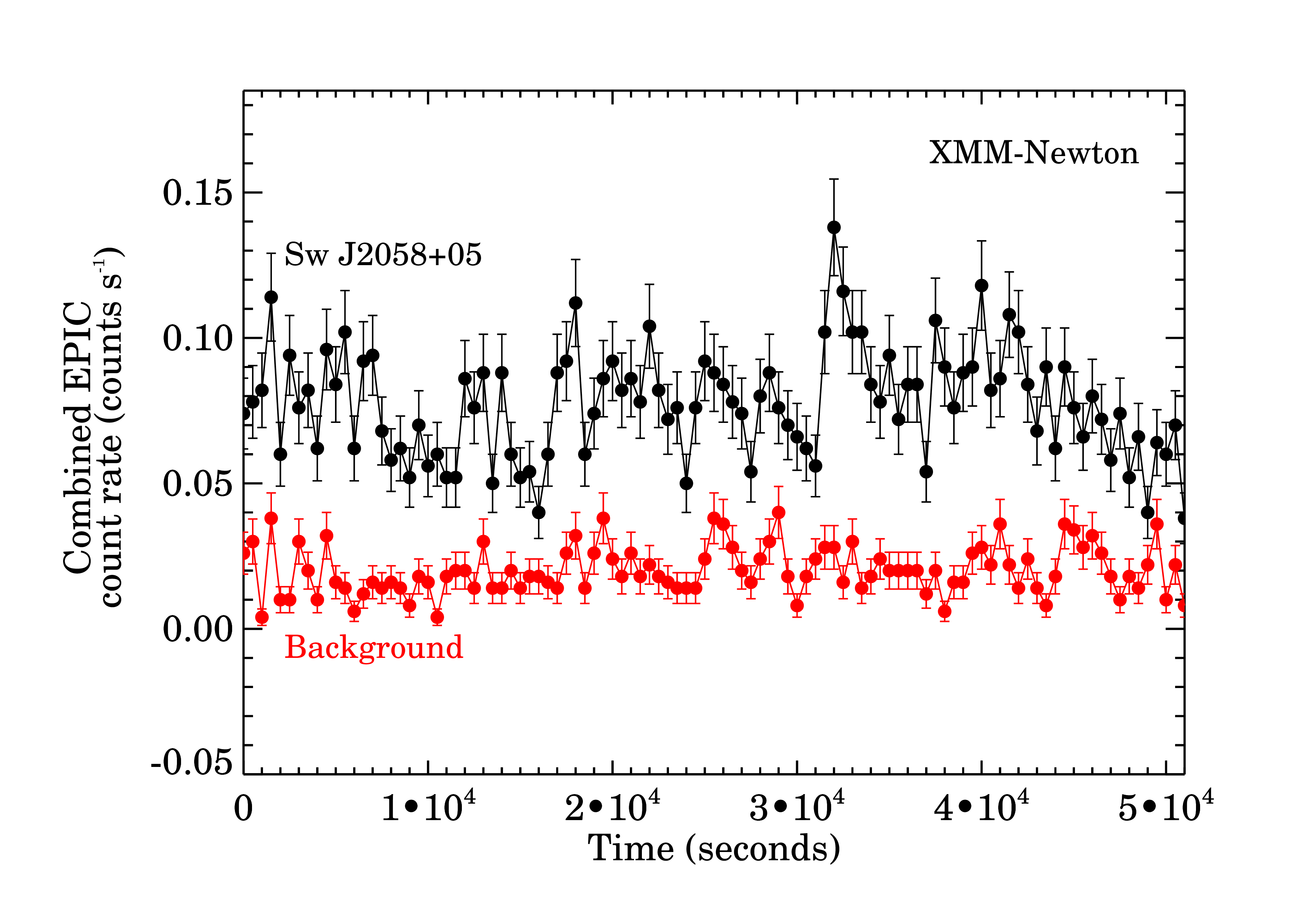}
\end{center}

{\textbf{Figure 2:} {\it XMM-Newton}/EPIC (both pn and MOS) X-ray (0.3--10\,keV) light 
curve of Sw~J2058+05 (filled black circles), highlighting X-ray variability on timescales 
of $\sim 500$\,s. The light curve was derived from the longest good 
time interval of 50\,ks from the {\it XMM-Newton} observation with ID 0694830201 and binned 
at 500\,s. The background during the same time is shown in red. The source 
light curve was fit to a constant-flux model and shows clear variability ($\chi^2 = 236$ for 102 
degrees of freedom). }
\label{fig:figure2}
\end{figure}
\clearpage

\begin{figure}

\begin{center}
\includegraphics[width=5.5in, height=4.25in, angle=0]{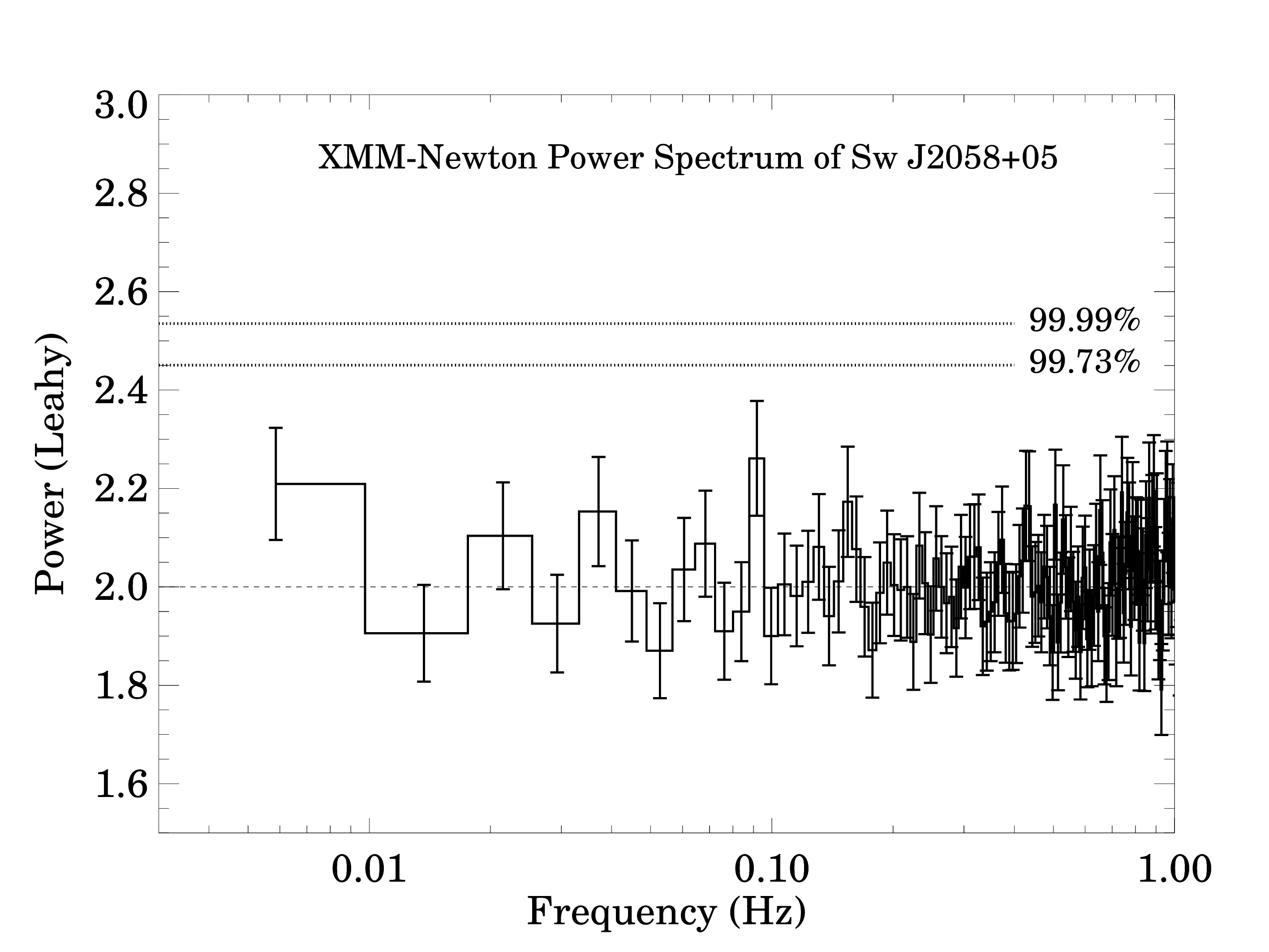}
\end{center}

{\textbf{Figure 3:} {\it XMM-Newton}/EPIC-pn (ObsID: 0694830201) X-ray (0.3--10\,keV) power density spectrum of 
Sw~J2058+05. The power spectrum is Leahy normalized (Leahy 1983) with a Poisson noise level of 2 (dashed horizontal line). The frequency 
resolution is 7.8\,mHz and each bin is an average of 188 independent power spectral measurements. The confidence 
limits ($3\sigma$/99.73\% and $3.9\sigma$/99.99\%) are indicated by the two horizontal dotted lines. The spectrum 
is featureless and consistent with being flat (white noise).   }
\label{fig:figure3}
\end{figure}
\clearpage


\begin{figure}

\begin{center}
\includegraphics[width=6.5in, height=4.15in, angle=0]{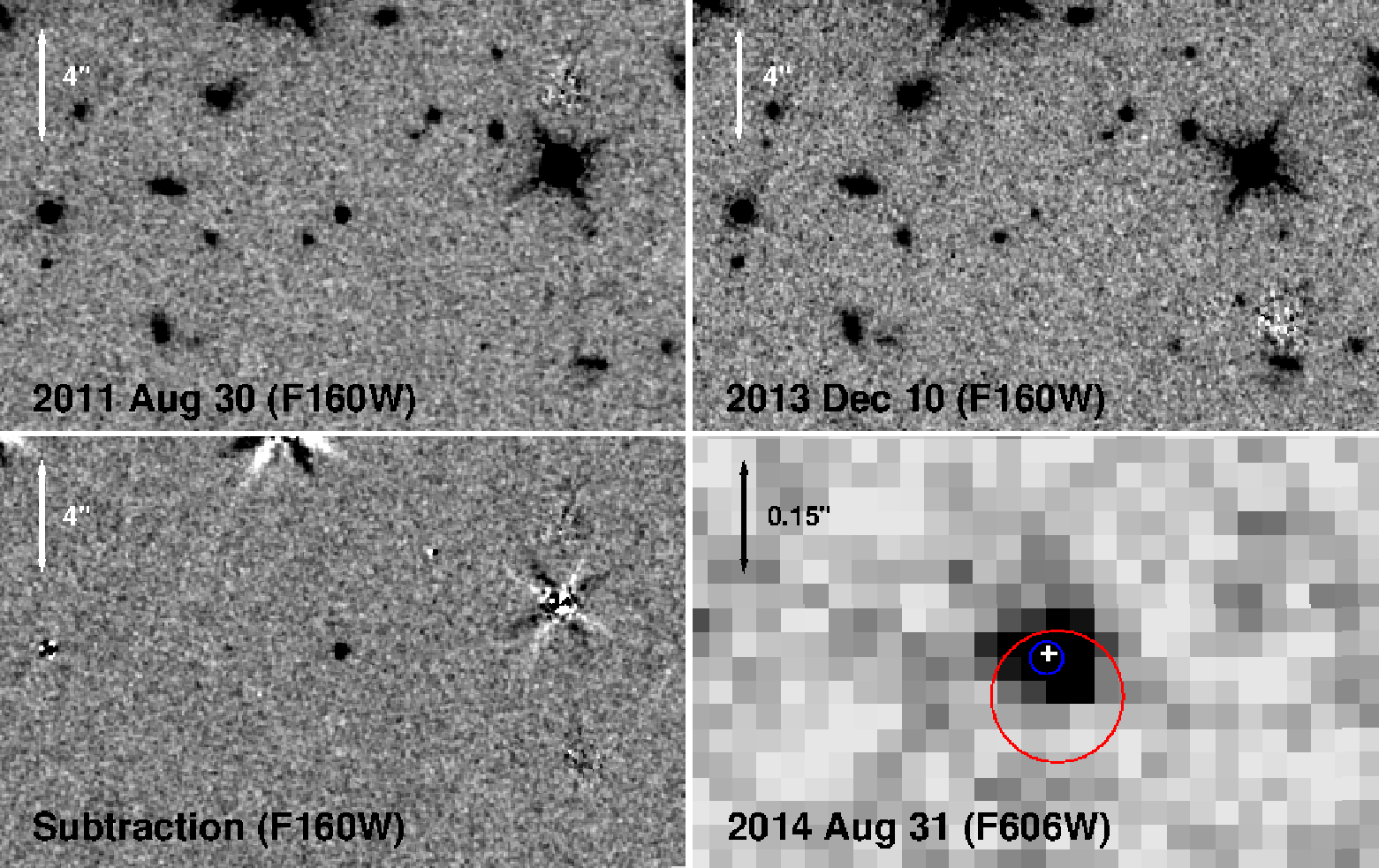}
\end{center}
{\textbf{Figure 4:} {\it Top-left panel:} {\it HST}/WFC3 $F160W$ image of the location of Sw~J2058+05 obtained 
on 2011 Aug. 30 (two months after Sw~J2058+05 reached its peak luminosity). {\it 
Top-right panel:} An image of Sw~J2058+05 with the identical instrument configuration from 2013 Dec. 10 
(long after the outburst when the optical emission is dominated by the host galaxy). {\it Bottom-left panel: } 
Digital image subtraction of the two $F160W$ frames. To within measurement uncertainties, the location of the 
resulting transient emission is consistent with the centroid of the host galaxy. {\it Bottom-right panel: } 
Zoomed-in image of the host galaxy in the $F606W$ filter. The centroid of the host galaxy is indicated by the 
white cross. Our most precise astrometric constraints come from aligning this $F606W$ image from 2014 Aug 31 
with a previous HST image of the transient from 2011 Aug 30 in the $F475W$ filter, for which the 68\% confidence 
uncertainty in the astrometric tie between the two frames is 23 mas in radius (blue circle). The VLBA position 
for Sw~J2058+05, along with the uncertainty in connecting the VLBA position to the HST astrometric frame (68\% 
confidence radius of 90 mas) is indicated by the red circle.
}
\label{fig:figure4}
\end{figure}
\clearpage

\begin{figure}

\begin{center}
\includegraphics[width=6.in, height=5.85in, angle=0]{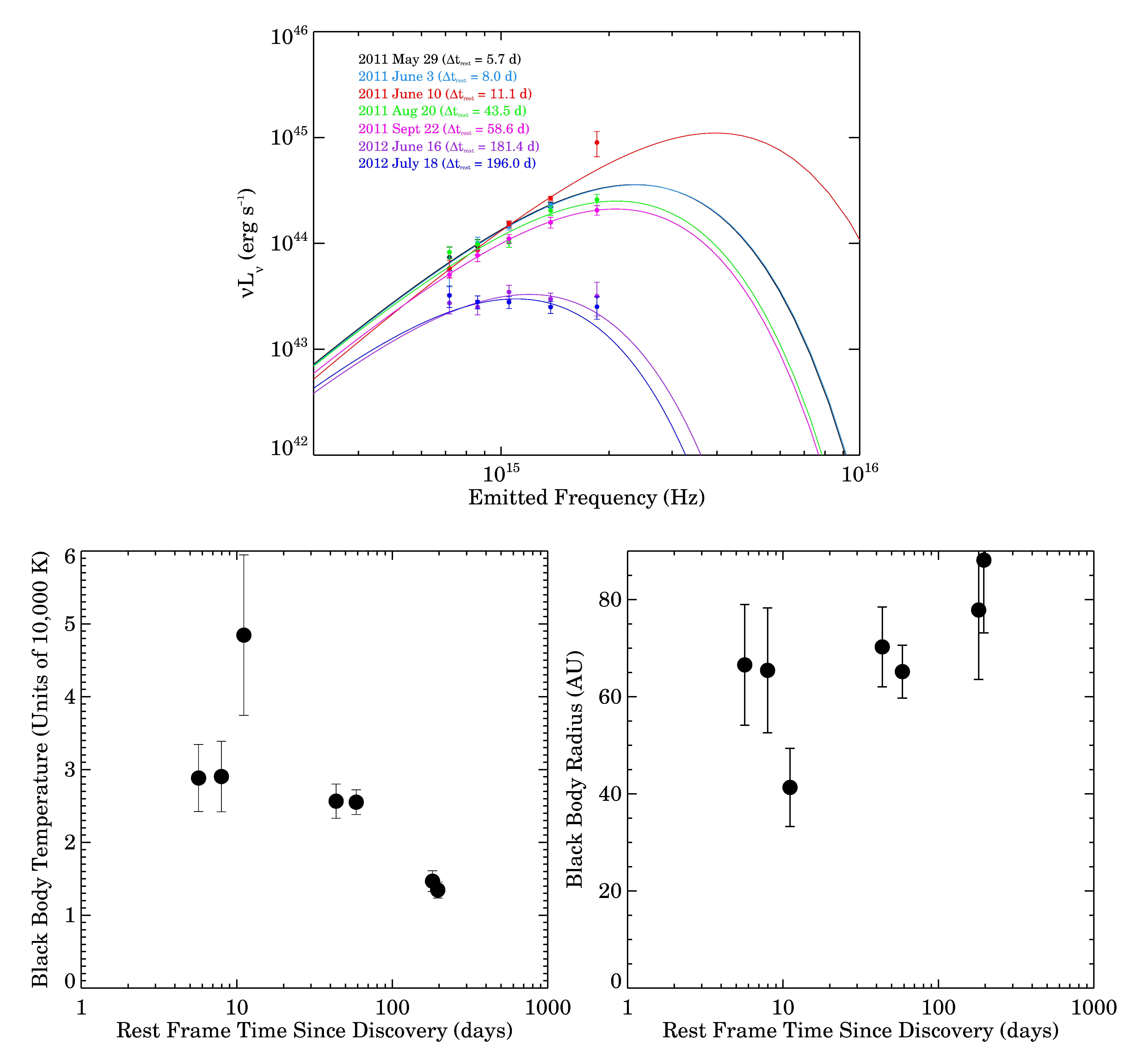}
\end{center}

{\textbf{Figure 5:} {\it Top panel:} UVOIR SEDs of Sw~J2058+05 at various epochs (filled circles). 
The best-fit single-temperature blackbodies (solid curves) are also shown. $\Delta$t$_{rest}$ refers to 
days in rest frame since discovery. {\it Bottom-left panel:} 
The blackbody temperature as a function of the rest-frame time since discovery. {\it Bottom-right panel:} The blackbody radius as a function of the rest-frame time since discovery. All of the 
error bars indicate 90\% confidence limits. }
\label{fig:figure5}
\end{figure}
\clearpage

\begin{figure}

\begin{center}
\includegraphics[width=6.in, height=4.85in, angle=0]{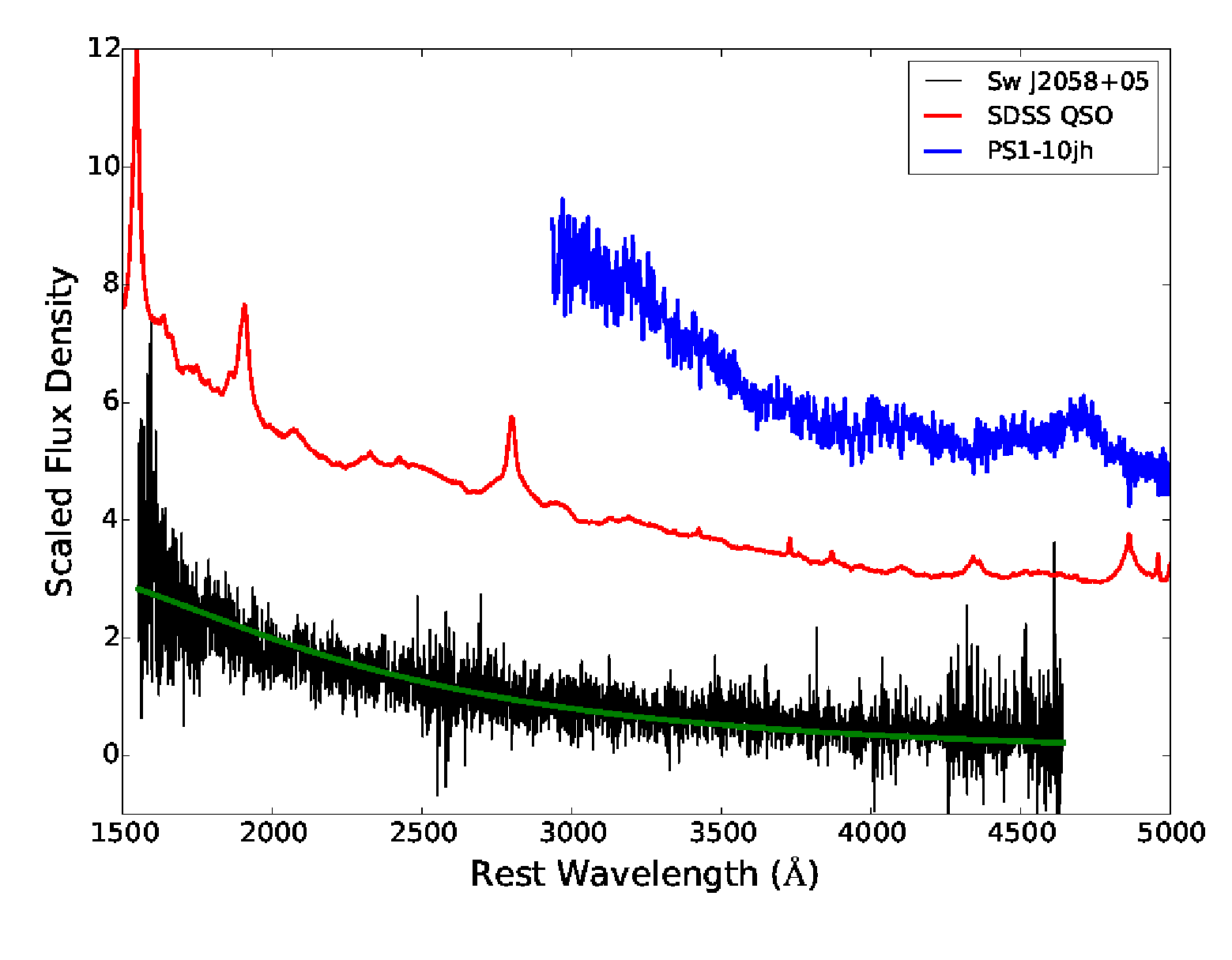}
\end{center}

{\textbf{Figure 6:} Optical and NIR spectra of Sw~J2058+05 (black) taken with Keck/LRIS on 2011 Aug. 28 
($\sim 47$\,d after discovery, measured in the rest frame).  The solid green line shows a fit
to a single blackbody with $T_{\mathrm{BB}} = (2.3 \pm 0.1) \times 10^{4}$\,K.
For comparison, the composite SDSS spectrum of 
quasars (Vanden Berk et al. 2001) and the spectrum of the TDE PS 1-10jh taken at its 
peak luminosity (Gezari et al. 2012) are shown in red and blue, respectively. The spectrum of 
Sw~J2058+05 does not contain any apparent absorption or emission lines at this stage.}
\label{fig:figure6}
\end{figure}
\clearpage

\begin{table}
\caption{Summary of X-ray Spectral Modeling of Sw J2058+05}
\begin{center}
{\footnotesize
\begin{tabular}{ccccc}
\toprule
\toprule \\
  {\bf Instrument}  		& {\bf ObsID}		& {\bf MJD Date}$\dagger$$\dagger$  & {\bf X-ray Flux$^{\ast}$}  		& {\bf Notes}$\dagger$ \\ 
\\
\midrule
{\it Swift}/XRT    	&     00032004001    	&           55708.915    &           48.12$^{+1.61}_{-1.70}$    &    PC Mode \\
{\it Swift}/XRT    	&     00032004002    	&           55711.582    &           64.49$^{+2.20}_{-2.19}$    &    WT Mode \\
{\it Swift}/XRT    	&     00032004003    	&           55714.412    &           59.37$^{+2.24}_{-2.08}$    &    WT Mode \\
{\it Swift}/XRT    	&     00032004004    	&           55717.879    &           48.78$^{+1.62}_{-1.51}$    &    WT Mode \\
{\it Swift}/XRT    	&     00032004005    	&           55720.568    &           31.78$^{+1.66}_{-1.59}$    &    WT Mode \\
{\it Swift}/XRT    	&     00032004007    	&           55726.045    &           12.61$^{+0.90}_{-0.88}$    &    WT Mode \\
{\it Swift}/XRT    	&     00032004008    	&           55729.110    &           15.17$^{+1.08}_{-1.02}$    &    WT Mode \\
{\it Swift}/XRT    	&     00032004009    	&           55735.539    &           9.58$^{+0.84}_{-0.78}$    &    WT Mode \\
{\it Swift}/XRT    	&     00032004010    	&           55738.548    &           8.01$^{+0.79}_{-0.89}$    &    WT Mode \\
{\it Swift}/XRT    	&     00032026001    	&           55743.760    &           3.33$^{+0.74}_{-0.64}$    &    WT Mode \\
{\it Swift}/XRT    	&     00032026002    	&           55748.457    &           2.85$^{+0.38}_{-0.43}$    &    WT Mode \\
{\it Swift}/XRT    	&     00032026003    	&           55753.531    &           1.59$^{+0.38}_{-0.34}$    &    PC Mode \\
{\it Swift}/XRT    	&     00032026004    	&           55760.699    &           1.90$^{+0.36}_{-0.34}$    &    PC Mode \\
{\it Swift}/XRT    	&     00032026005    	&           55763.907    &           2.69$^{+0.64}_{-0.64}$    &    PC Mode \\
{\it Swift}/XRT    	&     00032026006    	&           55768.723    &          0.99$^{+0.25}_{-0.23}$    &    PC Mode \\
{\it Swift}/XRT    	&     00032026007    	&           55773.203    &           1.30$^{+0.28}_{-0.32}$    &    PC Mode \\
{\it Swift}/XRT    	&     00032026009    	&           55783.373    &           1.05$^{+0.24}_{-0.24}$    &    PC Mode \\
{\it Swift}/XRT    	&     00032026010-012	&           55806.229    &          0.64$^{+0.17}_{-0.15}$    &    PC Mode \\
{\it Swift}/XRT    	&     00032026013-015   &           55853.296    &          0.37$^{+0.09}_{-0.11}$    &    PC Mode \\
{\it Swift}/XRT    	&     00032026016-021   &           55885.110    &          0.14$^{+0.05}_{-0.05}$    &    PC Mode \\
{\it XMM-Newton}/EPIC   &     0679380801    	&           55885.635    &          0.19$^{+0.02}_{-0.02}$    &    Exposure: 23 ks \\
{\it XMM-Newton}/EPIC   &     0679380901    	&           55887.787    &          0.16$^{+0.02}_{-0.02}$    &    Exposure: 29 ks \\
{\it XMM-Newton}/EPIC   &     0694830201    	&           56049.048    &          0.17$^{+0.01}_{-0.01}$    &    Exposure: 55 ks \\
{\it Chandra}/ACIS	& 14975 		& 	    56383.806 	 & 	 $\le$ 1.05$\times$10$^{-3}$ 	  	& Exposure: 30 ks\\
{\it Chandra}/ACIS	& 16498			& 	    56594.972 	 &    $\le$ 1.76$\times$10$^{-3}$ 		& Exposure: 20 ks\\
{\it Chandra}/ACIS	& 14976 		& 	    56597.639 	 &    $\le$ 1.63$\times$10$^{-3}$ 	  	& Exposure: 30 ks\\
\bottomrule
\bottomrule \\
\end{tabular}\\
{\textsuperscript{$\dagger$$\dagger$}The source was discovered on MJD 55698. \textsuperscript{$\ast$}The X-ray fluxes were estimated in the bandpass of 0.3--10\,keV and have units of $10^{-12}$\,erg\,s$^{-1}$\,cm$^{-2}$. These represent the values just outside our Galaxy. The X-ray luminosities in the top panel of Figure 1 were estimated as flux $\times 4\pi D^{2}$, where $D$ is 8200\,Mpc. See text for details on the modeling. \textsuperscript{$\dagger$}PC refers to photon counting and WT to windowed timing. }
}
\end{center}
\end{table}
\clearpage

\begin{table}
\begin{center}
\caption{A Summary of UV/Optical/IR observations of Sw J2058+05}

{\scriptsize
\begin{tabular}{cccccc}
\toprule
\toprule \\
  {\bf UTC} 	& {\bf MJD Date}  & {\bf Telescope} 	& {\bf Filter}  & {\bf Exposure}  	& {\bf AB Magnitude$^{\ast}$}   \\ 
  {\bf date} 	&   		  &  			&   		& {\bf (seconds)} 	&    \\ 
\\
\midrule 
2011 Aug 12.05 	& 55785.05 	  & VLT - HAWK-I 	& J 				& 1020  		& $22.72 \pm 0.33 $ \\
2011 Aug 12.07  & 55785.07 	  & VLT - HAWK-I 	& K 				& 1080 			&  $>21.6$  \\
2011 Aug 20.07 	& 55793.07 	  & VLT - FORS 2 	& u 				& 840.0 		& $22.86 \pm 0.13$  \\
2011 Aug 20.07  & 55793.07 	  & VLT - FORS 2 	& g 				& 120.0 		& $22.69 \pm 0.11$  \\
2011 Aug 20.07  & 55793.07 	  & VLT - FORS 2 	& r 				& 120.0 		& $23.07 \pm 0.11$  \\
2011 Aug 20.08  & 55793.08 	  & VLT - FORS 2 	& i 				& 200.0 		& $22.84 \pm 0.11$  \\
2011 Aug 20.08 	& 55793.08 	  & VLT - FORS 2 	& z 				& 720.0 		& $22.79 \pm 0.14$  \\
2011 Aug 30.56  & 55803.56  	  & HST - WFC3  	& F160W (H band)$^{\dagger\dagger}$   			& 1196.9   		& $23.36 \pm 0.02$  \\
2011 Aug 30.58  & 55803.58  	  & HST - WFC3  	& F475W (SDSS g)$^{\dagger}$ 	& 1110.0   		& $23.06 \pm 0.02$  \\
2011 Sept 2.21  & 55806.21 	  & VLT - HAWK-I 	& J 				& 1020 			& $ 22.22 \pm 0.22 $ \\
2011 Sept 2.21  & 55806.21 	  & VLT - HAWK-I 	& K 				& 1080 			& $ 21.87 \pm 0.25 $  \\
2011 Sept 22.06 & 55826.06 	  & VLT - FORS 2 	& r 				& 120.0 		& $22.98 \pm 0.10 $  \\
2011 Sept 22.07 & 55826.07 	  & VLT - FORS 2 	& u 				& 840.0 		& $23.11 \pm 0.11$  \\
2011 Sept 22.07 & 55826.07 	  & VLT - FORS 2 	& g 				& 120.0 		& $22.97 \pm 0.13$ \\
2011 Sept 22.08 & 55826.07 	  & VLT - FORS 2 	& i 				& 200.0 		& $23.10 \pm 0.14$   \\
2011 Sept 22.08 & 55826.07 	  & VLT - FORS 2 	& z 				& 720.0 		& $23.32 \pm 0.08$   \\
2011 Sept 24.99 & 55828.99 	  & VLT - HAWK-I 	& J 				& 1020 			& $22.46 \pm 0.16 $ \\
2011 Sept 25.01 & 55829.01 	  & VLT - HAWK-I 	& K 				& 1080  		& $21.60 \pm 0.20 $ \\
2011 Nov 20.02  & 55885.02 	  & Gemini-S  - GMOS 	& u 				& 300.5 		& $>24.5$ \\
2011 Nov 20.02  & 55885.02 	  & Gemini-S - GMOS 	& g 				& 100.5 		& $23.93 \pm 0.21 $  \\
2011 Nov 20.03  & 55885.02 	  & Gemini-S - GMOS 	& r 				& 100.5 		& $23.53 \pm 0.16 $  \\
2011 Nov 30.96  & 55894.96  	  & HST - WFC3  	& F160W (H band)$^{\dagger\dagger}$  			& 1196.9   		& $23.56 \pm 0.02$  \\ 
2011 Nov 30.99  & 55894.99  	  & HST - WFC3  	& F475W (SDSS g)$^{\dagger}$	& 1110.0   		& $23.89 \pm 0.02$  \\
2012 June 16.32 & 56094.32 	  & VLT - FORS 2 	& r 				& 400.0 		& $24.24 \pm 0.17 $ \\
2012 June 16.33 & 56094.33 	  & VLT - FORS 2 	& g 				& 400.0 		& $24.78 \pm 0.15$  \\
2012 June 16.34 & 56094.34 	  & VLT - FORS 2 	& u  				& 840.0 		& $25.14 \pm 0.38$  \\
2012 June 16.35 & 56094.35 	  & VLT - FORS 2 	& i 				& 240.0 		& $24.32 \pm 0.17$  \\
2012 June 16.35 & 56094.35 	  & VLT - FORS 2 	& z 				& 720.0 		& $23.99 \pm 0.23$  \\
2012 July 18.27 & 56126.27 	  & VLT - FORS2 	& u 				& 840 			& 25.39 $\pm$ 0.26\\
2012 July 18.26  & 56126.26 	  & VLT - FORS2 	& g 				& 400 			& 24.97 $\pm$ 0.14\\
2012 July 18.26  & 56126.26 	  & VLT - FORS2 	& r 				& 400  			&  24.48 $\pm$ 0.14\\
2012 July 18.28  & 56126.28 	  &  VLT - FORS2 	& i  				& 240 			& 24.20 $\pm$ 0.15 \\
2012 July 18.29  & 56126.29 	  & VLT - FORS2 	& z  				& 720 			& 23.81 $\pm$ 0.25 \\
2012 Aug 22.09 & 56161.09 	  & VLT - FORS2 	& u  				& 840  			& $>25.8$ \\
2012 Aug 22.08 & 56161.08 	  & VLT - FORS2 	& g  				& 400  			& $>26.4$\\
2012 Aug 22.08 & 56161.08 	  & VLT - FORS2 	& r  				& 400  			& $>26.0$\\
2012 Aug 22.10 & 56161.10 	  & VLT - FORS2 	& i   				& 240 			& $>24.9$ \\
2012 Aug 22.10 & 56161 		  & VLT - FORS2 	& z  				& 720  			& $>25.2$\\
2012 Oct 09.01 & 56209.01 	  & VLT - FORS2 	& u  				& 840 			&$>26.0$  \\
2012 Oct 09.01 & 56209.01 	  & VLT - FORS2 	& g  				& 400 			& $>26.3$ \\
2012 Oct 09.00 & 56209.00 	  & VLT - FORS2 	& r  				& 400  			& $>26.2$ \\
2012 Oct 09.02 & 56209.02 	  & VLT - FORS2 	& i   				& 240 			& $>24.8$ \\
2012 Oct 09.03 & 56209.03 	  & VLT - FORS2 	& z  				& 720 			& $>25.2$ \\
2013 Dec 10.58  & 56636.58  	  & HST - WFC3  	& F160W (H band)$^{\dagger\dagger}$   			& 2611.8   		& $25.99 \pm 0.08$ \\
2014 Aug 31.48  & 56900.48  	  & HST/ACS - WFC   	& F606W   			& 5236.0   		& $26.78 \pm 0.10$ \\
\bottomrule	
\bottomrule \\
\end{tabular}\\
{\textsuperscript{$\ast$}Reported magnitudes have not been corrected for Galactic extinction (E(B - V) = 0.095 mag; Schlafly \& Finkbeiner 2011). Upper limits represent 3$\sigma$ uncertainties. \textsuperscript{$\dagger$}{\it HST/F475W} filter has a bandpass similar to SDSS's g band. \textsuperscript{$\dagger\dagger$}{\it HST/F160W} filter has a bandpass similar to the standard H band.}
}
\end{center}
\end{table}
\clearpage

\begin{table}
\begin{center}
\caption{Optical/Near-IR Spectra of Sw J2058+05}

\begin{tabular}{ccccc}
\toprule
\toprule \\
{\bf Date} 		& {\bf Telescope/Instrument} 	& {\bf Wavelength} & {\bf Exposure} 	   & {\bf SNR$^{\ast}$} \\
{\bf (UT)} 		& 				& {\bf (\AA)}     & {\bf (s)} 	   &		   \\ 
\\
\midrule
2011 Aug 2.41 		& Keck/LRIS (blue) 		& 3360--5600 	  & 1800.0 	  	   & 3.4 \\
2011 Aug 2.41 		& Keck/LRIS (red) 		& 5600--10,200     & 1800.0 		   & 2.0 \\
2011 Aug 4.16 		& VLT/FORS 			& 3400--6100 	      & 4800.0 		   & 2.4 \\
2011 Aug 28.47 		& Keck/LRIS (blue) 		& 3360--5600      & 1800.0 	 	   & 5.3 \\
2011 Aug 28.47 		& Keck/LRIS (red) 		& 5600--10,200     & 1800.0 		   & 2.9 \\
2011 Sep 2.04 		& VLT/XSHOOTER (UV) 		& 3000--5560   & 3600.0 		   & 0.5 \\
2011 Sep 2.04 		& VLT/XSHOOTER (VIS) 		& 5300--10,200  & 3600.0 		   & 0.2 \\
2011 Sep 2.04 		& VLT/XSHOOTER (NIR) 		& 9900--24,800  & 3600.0 		   & 0.1 \\
\bottomrule
\bottomrule \\
\end{tabular}\\
{\textsuperscript{$\ast$}Per resolution element.}
\end{center}
\end{table}
\clearpage


\begin{table}
\begin{center}
\caption{Summary of {\it XMM-Newton} X-ray (0.3--10\,keV) Spectral Modeling of Sw J2058+05}

\begin{tabular}{cccccc}

\toprule
\toprule \\
  {\bf ObsID$^{a}$}	& {\bf Absorbing}  	 & {\bf Power-law} 		& {\bf Power-law} 	& {\bf $\chi^2$/dof}		& {\bf X-ray Flux$^{d}$}   \\ 
 			& {\bf column$^{b}$}     & {\bf index$^{c}$}		& {\bf Normalization}   &				& 		     \\ 
\\
\midrule 
 0679380801		& 0.23$^{+0.15}_{-0.13}$ & 1.89$^{+0.15}_{-0.13}$  	& 3.6$^{+0.4}_{-0.4}$ 	& 53/48				& 0.19$^{+0.02}_{-0.02}$     \\  
 0679380901		& 0.15$^{+0.18}_{-0.16}$ & 1.81$^{+0.18}_{-0.16}$    	& 2.8$^{+0.4}_{-0.3}$   & 55/64  			& 0.16$^{+0.02}_{-0.02}$ \\  
 0694830201		& 0.19$^{+0.13}_{-0.12}$ & 1.67$^{+0.10}_{-0.10}$    	& 2.5$^{+0.2}_{-0.2}$   & 86/94				& 0.17$^{+0.01}_{-0.01}$ \\  
\bottomrule
\bottomrule \\
\end{tabular}\\
{\textsuperscript{a}{\it XMM-Newton} assigned observation ID. \textsuperscript{b}Units of $10^{22}$\,atoms\,cm$^{-2}$. 
\textsuperscript{c}The X-ray spectra were modeled with {\it phabs$*$zwabs$*$pow} in XSPEC. The Galactic column 
({\it phabs}) was fixed at $0.088 \times 10^{22}$\,atoms\,cm$^{-2}$ (Kalberla et al. 2005 \& Willingale et al. 2013) and 
the redshift in {\it zwabs} was fixed at 1.1853 (C12). \textsuperscript{d}The X-ray fluxes were 
estimated in the bandpass of 0.3--10\,keV and have units of $10^{-12}$\,erg\,s$^{-1}$\,cm$^{-2}$. }
\end{center}
\end{table}
\clearpage

\begin{table}
\begin{center}
\caption{Summary of UVOIR SED Modeling of Sw J2058+05$^{a}$}

\begin{tabular}{ccccc}
\toprule
\toprule \\
  {\bf UTC}		& {\bf MJD Date}  	 		 & {\bf Blackbody} 			& {\bf Blackbody } 		& {\bf $\chi^2$/dof}		\\ 
 			& {\bf (rest-frame days since discovery)}    	 & {\bf temperature$^{b}$}		& {\bf radius$^{c}$}   		&				\\ 
\\
\midrule 
 2011 May 29		& 55710.41				 & 2.9$\pm$0.5  			& 66.6$\pm$12.4 		& 0.3/2     \\
			& (5.67)				 &					&			 	&				\\
 2011 June 3		& 55715.40				 & 2.9$\pm$0.5  			& 65.4$\pm$12.9 		& 0.3/2     \\
			& (7.96)				 &					&			 	&				\\
 2011 June 10		& 55722.26				 & 4.9$\pm$1.1  			& 41.3$\pm$8.1 			& 3.8/3     \\
			& (11.10)				 &					&			 	&				\\
 2011 Aug 20		& 55793.07				 & 2.6$\pm$0.2  			& 70.3$\pm$8.2 			& 12.3/3     \\
			& (43.51)				 &					&			 	&				\\
 2011 Sept 22		& 55826.07				 & 2.6$\pm$0.2  			& 65.2$\pm$5.4 			& 0.2/3     \\
			& (58.60)				 &					&			 	&				\\
 2012 June 16		& 56094.34				 & 1.5$\pm$0.2  			& 71.3$\pm$15.5 		& 2.5/3     \\
			& (181.37)				 &					&			 	&				\\
 2012 July 18		& 56126.27				 & 1.4$\pm$0.1  			& 88.1$\pm$15.0 		& 4.7/3     \\
			& (196.0)				 &					&			 	&				\\
 \bottomrule
\bottomrule \\
\end{tabular}\\
{\textsuperscript{a}Data for the first three epochs was acquired by C12 while the rest are from Table 2. \textsuperscript{b}Units of 10,000\,K. 
\textsuperscript{c}Units of AU (astronomical unit). The SEDs were modeled with a single-temperature blackbody. }
\end{center}
\end{table}
\clearpage

\vfill\eject

\vfill



\end{document}